\documentclass[aps,prd,nofootinbib,amsmath,superscriptaddress]{revtex4}

\usepackage{bm}
\usepackage{graphicx}
\usepackage{caption}
\usepackage{subcaption}
\newcommand{\de}{{\rm d}}
\newcommand{\be}{\begin{equation}}
\newcommand{\ee}{\end{equation}}
\newcommand{\bea}{\begin{eqnarray}}
\newcommand{\eea}{\end{eqnarray}}

\newcommand{\bfk}{\mbox{\boldmath $k$}}
\def\bkt{\bfk_\perp}
\def\kt{k_\perp}
\newcommand{\bfp}{\mbox{\boldmath $p$}}

\def\bpp{\bfp_\perp}
\newcommand{\bfq}{\mbox{\boldmath $q$}}
\newcommand{\bfP}{\mbox{\boldmath $P$}}

\def\pp{p_\perp}

\def\avp{\langle p_\perp ^2\rangle}

\def\avPT{\langle P_T^2\rangle}

\begin{document}

\title{On the extraction of the valence transversity distributions from
SIDIS data}

\author{M.~Anselmino}
\email{mauro.anselmino@to.infn.it}
\affiliation{Dipartimento di Fisica, Universit\`a di Torino,
             Via P.~Giuria 1, I-10125 Torino, Italy}
\affiliation{INFN, Sezione di Torino, Via P.~Giuria 1, I-10125 Torino, Italy}
%
\author{R.~Kishore}
\email{raj.theps@gmail.com}
\affiliation{Department of Physics, Indian Institute of Technology Bombay,
Mumbai-400076, India}
\author{A.~Mukherjee}
\email{asmita@phy.iitb.ac.in}
\affiliation{Department of Physics, Indian Institute of Technology Bombay,
Mumbai-400076, India}

\date{\today}

\begin{abstract}
The transversity distribution for $u$ and $d$ quarks is usually extracted 
from data on spin asymmetries in Semi Inclusive Deep Inelastic Scattering (SIDIS): 
however, due to its chiral odd nature, it has to be coupled to another chiral odd 
function, typically the Collins or the di-hadron fragmentation function. A recent 
suggestion of considering SIDIS data involving ratios of spin asymmetries and
avoiding a knowledge of the Collins function, is briefly discussed. New 
measurements, involving ratios of cross sections, are suggested. They would allow 
a direct extraction of the transversity ratio, $h_1^{d_V}/h_1^{u_V}$. Numerical 
estimates are given.               
\end{abstract}


\maketitle

\section{\label{intro}Introduction and formalism}
The quark transversity distribution, $h_1^q(x)$ or $\Delta_T^q(x)$, is the 
least known of the three basic parton distributions which describe the 
1-dimensional collinear representation of the partonic nucleon structure
(for a recent review paper, see Ref.~\cite{Anselmino:2020vlp}).
It is of great interest, as its integral is related to the 
tensor charge, a fundamental quantity which can be computed in lattice 
QCD~\cite{Bhattacharya:2016zcn,Alexandrou:2017qyt,Gupta:2018lvp,
Yamanaka:2018uud,Lin:2020rut}. 
Because it is chiral odd, information on the transversity 
distribution can be obtained through observables which involve another chiral
odd function. This is usually done through spin asymmetries in SIDIS processes,
which are given by a convolution of the transversity distribution and the 
Transverse Momentum Dependent Fragmentation Function (TMD-FF) introduced by 
Collins~\cite{Collins:1992kk}. Indeed the first extraction of $\Delta_T^u(x)$ 
and $\Delta_T^d(x)$ was obtained in this 
way~\cite{Anselmino:2007fs,Anselmino:2008jk,Anselmino:2013vqa,Kang:2015msa}. 
Similar results were obtained by coupling the transversity distribution with a 
di-hadron fragmentation 
function~\cite{Bacchetta:2011ip,Bacchetta:2012ty,Radici:2015mwa}. 

Let us briefly recall the formalism adopted to extract the transversity
distributions from SIDIS data, through Collins asymmetries. Following 
Refs.~\cite{Bacchetta:2006tn,Anselmino:2011ch}, where all details can be found,
the differential cross section for the semi-inclusive production of a hadron $h$,
in the current fragmentation region, from the collision of an unpolarised
lepton beam off a transversely polarised target can be written, in the deeply
inelastic regime, as follows (see Eq.~(79) of Ref.~\cite{Anselmino:2011ch}):
\bea
\dfrac{\de\sigma^{\ell p(S_T)\to \ell^\prime h\,X}}
{\de x_{B}\, \de Q^2 \, \de z_h \, \de ^2\bm{P}_{T} \, \de \phi_S} &=&
\dfrac{2\alpha^2}{Q^4}\,\Bigl\{ \frac{1+(1-y)^2}{2}\,F_{UU} + \dots
\label{eq:dsig-sidis} \\
&+& \Bigl[ 
(1-y)\,\sin(\phi_h+\phi_S)\,F_{UT}^{\sin(\phi_h+\phi_S)} + \dots\,\Bigr]\Bigr\}\,,
\nonumber
\eea
where we have omitted all terms which do not contribute to the Collins 
asymmetry. $x_B$, $y$, $z_h$ and $Q$ are the usual SIDIS variables. The quark 
momentum inside the target (with momentum $\bfp$) is $\bfq = x\, \bfp + \bkt$,
while the momentum of the final hadron generated by the fragmentation of the 
scattered quark (with momentum $\bfq'$) is $\bfP_h = z\,\bfq' + \bpp$. 
Notice that, at order $k_\perp/Q$, $x_B = x$ and $z_h = z$.
$\bfP_T$ is the hadron transverse momentum in the $\gamma$*-nucleon c.m. frame 
and, again at order $k_\perp/Q$, is given by $\bfP_T = \bkt + z\,\bpp$.
$\phi_h$ and $\phi_S$ are, respectively, the azimuthal angle of the observed 
hadron and of the target polarisation vector {\it w.r.t.}~the
leptonic plane. The subscript $UT$ in the structure functions $F$
reminds that we are considering the case of an unpolarised lepton beam and a
transversely polarised nucleon target ($UU$ refers to the unpolarised situation).

In the SIDIS case, the asymmetries are often expressed through their azimuthal
moments,
\be
A_{UT}^{W(\phi_h,\phi_S)} = 2 \,
\frac{\int\,d\phi_h \, d\phi_S\,\left[ d\sigma^\uparrow -
d\sigma^\downarrow\right]\,W(\phi_h,\phi_S)}
{\int\,d\phi_h \, d\phi_S\,\left[ d\sigma^\uparrow + d\sigma^\downarrow\right]}\,,
\label{eq:azi-mom}
\ee
where $W(\phi_h,\phi_S)$ is the appropriate azimuthal weight function
required in order to isolate the specific contribution of interest and
$d\sigma^{\uparrow,\downarrow}$ is the differential cross section of
Eq.~(\ref{eq:dsig-sidis}) with $S_T = \> \uparrow,\downarrow$ denoting,
respectively, a transverse polarisation with azimuthal angle $\phi_S$ and
$\phi_S + \pi$. Then we simply have
\bea
d\sigma^\uparrow - d\sigma^\downarrow &=&
\frac{2\alpha^2}{Q^4}\,\Bigl\{ 
2(1-y)\,\sin(\phi_h + \phi_S)\,F_{UT}^{\sin(\phi_h+\phi_S)} + \dots \Bigr\}\,,
\label{eq:dsig-updown1}\\
d\sigma^\uparrow + d\sigma^\downarrow &=&
\frac{2\alpha^2}{Q^4}\,\Bigl\{ [1+(1-y)^2]\,F_{UU} + \dots \Bigr\}\,.
\label{eq:dsig-updown2}
\eea

As the Collins effect generates a $\sin(\phi_h + \phi_S)$ modulation, we find 
that the azimuthal moment (\ref{eq:azi-mom}) of the Collins asymmetry in SIDIS processes is given by, from Eqs.~(\ref{eq:dsig-updown1}) 
and~(\ref{eq:dsig-updown2}): 
\be
A_{UT}^{\sin(\phi_h + \phi_S)} =
\dfrac{2(1-y)\,F_{UT}^{\sin(\phi_h+\phi_S)}}
{\bigl[ 1+(1-y)^2\bigr]\,F_{UU}} \equiv D_{NN} 
\dfrac{F_{UT}^{\sin(\phi_h+\phi_S)}}{F_{UU}}
\label{eq:AUT-coll}
\ee
where $D_{NN} = 2(1-y)/[1+(1-y)^2]$ is the quark depolarisation factor. 
$F_{UU}$ can be expressed as a convolution (meaning $\bkt$ and $\bpp$ 
integrations) of transverse momentum dependent unpolarised distribution
and fragmentation functions, while $F_{UT}^{\sin(\phi_h+\phi_S)}$ is a 
convolution of transversity distributions and Collins fragmentation functions
(precise expressions can be found, for example, in Eqs. (63) and (75) of 
Ref.~\cite{Anselmino:2011ch}). $A_{UT}^{\sin(\phi_h + \phi_S)}$ is the quantity 
experimentally measured, which relates data to a combination of the 
unknown transversity and Collins functions. 

It is useful, for a better understanding and further use in Sections~\ref{new}
and~\ref{conc}, to give here explicit expressions of 
$F_{UU}$ and $F_{UT}^{\sin(\phi_h+\phi_S)}$, based on particular 
parameterisations of the various TMD involved (see Ref.~\cite{Anselmino:2018psi} 
for a definition of all variables and a complete collection of all details): 
\bea
F_{UU} &=& \sum_q\,e_q^2\,f_{q/p}(x)\,D_{h/q}(z)\,
\dfrac{e^{-P_T^2/\langle P_T^2\rangle}}{\pi\langle P_T^2\rangle}
\equiv \sum_q\,e_q^2\,f_{q/p}(x)\,D_{h/q}(z)\,A(P_T)
\label{eq:F-uu} \\
F_{UT}^{\sin(\phi_h + \phi_S)} &=&
\sum_q\,e_q^2\,h_1^q(x)\,\Delta^N D_{h/q^\uparrow}(z) \,
\sqrt{\frac{e}{2}}\,\frac{P_T}{M_C}\,\frac{\avp_C^2}{\avp}\,
\dfrac{e^{-P_T^2/\avPT_T}}{\pi\avPT_T^2}
\equiv \sum_q\,e_q^2\,h_1^q(x)\,\Delta^N D_{h/q^\uparrow}(z) \, B(P_T)
\label{eq:F-ut-coll}
\eea
where $f_{q/p}$ and $D_{h/q}$ are the usual collinear PDFs and FFs, while 
$\Delta^N D_{h/q^\uparrow}(z)$ is the $z$ dependent part of the Collins
functions. Notice that these expressions assume, as it is usually done in the 
phenomenological study of the transversity distributions from SIDIS data, a 
factorisation, in the TMDs, of the $x$, $z$, $\kt$ and $\pp$ dependences; 
the latter are taken to be Gaussian, and the $A(P_T)$ and $B(P_T)$ 
functions reflect their parameters. Different choices of the parameters and 
different forms of the factorised TMDs would not affect the general conclusions 
of the next Sections.          
         
\section{\label{new}Suggested measurements} 

In a recent paper~\cite{Barone:2019yvn} a suggestion was made which could avoid, 
in the extraction of the transversity distributions $h_1^q(x)$ from SIDIS data, 
a dependence on the Collins functions. It simply amounts to introduce particular 
asymmetries involving combinations of cross sections measured for different targets 
and different final hadrons; in these observables the Collins functions cancel out 
and one remains with ratios of transversity distributions, in addition to 
unpolarised PDFs and FFs. 

In order to describe the suggestion of Ref.~\cite{Barone:2019yvn}, let us follow
their notations and rewrite Eq.~(\ref{eq:dsig-sidis}) as:
\be
\sigma_t^{\pm} = \sigma_{0,t}^{\pm} + \sin(\phi_h+\phi_S)\,
D_{NN} \, \sigma_{C,t}^{\pm} 
+ \dots \label{csts}
\ee 
where the subscript $t$ indicates the kind of target ($p$ for proton, $n$ for 
neutron and $d$ for deuteron) and the superscript $+$ or $-$ refers to positive 
or negative pions. By comparing Eqs.~(\ref{eq:dsig-sidis}) and~(\ref{csts}),
one can easily extract the expressions of $\sigma_0$ and $\sigma_C$: 
\be
\sigma_0 = \frac{\alpha^2}{Q^4} \, \bigl[ 1+(1-y)^2\bigr]\,F_{UU}
\quad\quad\quad
\sigma_C = \frac{\alpha^2}{Q^4} \, \bigl[ 1+(1-y)^2\bigr] \,
F_{UT}^{\sin(\phi_h + \phi_S)} \>.
\ee
In addition, in Ref.~\cite{Barone:2019yvn}, it is taken as a measure of the 
Collins asymmetry the ratio
\be 
A_C = \frac{\sigma_C}{\sigma_0} = \dfrac{F_{UT}^{\sin(\phi_h + \phi_S)}}{F_{UU}}
= \frac{1}{D_{NN}} \, A_{UT}^{\sin(\phi_h + \phi_S)} \>,
\ee    
which only differs by the usually measured asymmetry by the $1/D_{NN}$ factor.

Using the expressions for $F_{UU}$ and $F_{UT}^{\sin(\phi_h + \phi_S)}$
given in Eqs.~(\ref{eq:F-uu}) and~(\ref{eq:F-ut-coll}) (considering, for the 
moment, the case of a generic final hadron $h$ and a proton target) one has
\bea
\sigma_{C,p}^h &=& \frac{\alpha^2}{Q^4} \, \bigl[ 1+(1-y)^2\bigr] \,
\sum_q e_q^2 \, h_1^q(x) \, \Delta^N D_{h/q^\uparrow}(z) \, B(P_T) \label{scp}\\
\sigma_{0,p}^h &=& \frac{\alpha^2}{Q^4} \, \bigl[ 1+(1-y)^2\bigr] \,
\sum_q e_q^2 \, f_{q/p}(x) \, D_{h/q}(z) \, A(P_T) \>. \label{s0p}
\eea

Following a previous similar suggestion for helicity 
distributions~\cite{Christova:2000nz} in Ref.~\cite{Barone:2019yvn} it is 
defined a difference asymmetry as 
\be
A_{D,t} \equiv \dfrac{\sigma_{C,t}^+ - \sigma_{C,t}^-} 
{\sigma_{0,t}^+ + \sigma_{0,t}^-} = 
\dfrac{\sigma_{0,t}^+}{\sigma_{0,t}^+ + \sigma_{0,t}^-} \, A_{C,t}^+ - 
\dfrac{\sigma_{0,t}^-}{\sigma_{0,t}^+ + \sigma_{0,t}^-} \, A_{C,t}^- \>,
\label{adt}
\ee
where the second equality strictly holds only in case the Collins angle 
acceptance is the same for positive and negative particles. The quantity
given in Eq.~(\ref{adt}) can be obtained from available data on the Collins
symmetry and the unpolarised cross section. 

Assuming isospin symmetry and introducing favoured and disfavoured 
fragmentation and Collins functions:
\bea
&&f_{u/p}=f_{d/n}\equiv f_1^u \quad\quad f_{d/p}=f_{u/n}\equiv f_1^d \quad\quad
f_{\bar u/p}=f_{\bar d/n}\equiv f_1^{\bar u} \quad\quad 
f_{\bar d/p}=f_{\bar u/n}\equiv f_1^{\bar d} \\
&&f_{s/p}=f_{s/n}\equiv f_1^{s} \quad\quad 
f_{\bar s/p}=f_{\bar s/n}\equiv f_1^{\bar s} \\
&&D_{\pi^+/u}=D_{\pi^-/d}=D_{\pi^+/\bar d}=D_{\pi^-/\bar u} \equiv D_{1,fav} \\
&&D_{\pi^+/\bar u}=D_{\pi^-/\bar d}=D_{\pi^+/d}=D_{\pi^-/u} \equiv D_{1,dis}
\quad\quad D_{\pi^\pm/s,\bar s} \equiv D_{1,s} \\
&&\Delta^N D_{\pi^+/u^\uparrow}=\Delta^N D_{\pi^-/d^\uparrow} =
\Delta^N D_{\pi^+/\bar d^\uparrow}=\Delta^N D_{\pi^-/\bar u^\uparrow}
\equiv \Delta^N D_{fav} \\
&&\Delta^N D_{\pi^+/\bar u^\uparrow}=\Delta^N D_{\pi^-/\bar d^\uparrow} =
\Delta^N D_{\pi^+/d^\uparrow}=\Delta^N D_{\pi^-/u^\uparrow} \equiv \Delta^N D_{dis}
\quad\quad \Delta^N D_{\pi^\pm/s,\bar s} = \Delta^N D_{1, s} \>,
\eea
one can work out, from Eqs.~(\ref{scp}) and~(\ref{s0p}), the expressions of 
$\sigma_{C,t}^\pm$ and $\sigma_{0,t}^\pm$, as done in Ref.~\cite{Barone:2019yvn}. 
Dropping the common factor $(1/9)(\alpha^2/Q^4) \, [1+(1-y)^2] B(P_T)$, one has:
\bea
\sigma_{C,p}^+ &\sim& \left[ (4h_1^{u} + h_1^{\bar d})\,\Delta^N D_{fav} +
(4h_1^{\bar u} + h_1^{d})\,\Delta^N D_{dis}  
+ (h_1^{s} + h_1^{\bar s}) \, \Delta^N D_{1, s}\right] \label{scp+} \\
\sigma_{C,p}^- &\sim& \left[ (4h_1^{u} + h_1^{\bar d})\,\Delta^N D_{dis} +
(4h_1^{\bar u} + h_1^{d})\,\Delta^N D_{fav} 
+ (h_1^{s} + h_1^{\bar s}) \, \Delta^N D_{1, s} \right] \label{scp-} \\
\sigma_{C,d}^+ &\sim& \left[ (h_1^{u} + h_1^{d})\,(4\Delta^N D_{fav} + 
\Delta^N D_{dis}) + (h_1^{\bar u} + h_1^{\bar d})\,(\Delta^N D_{fav} + 
4\Delta^N D_{dis}) + 2(h_1^{s} + h_1^{\bar s}) \, \Delta^N D_{1, s}\right] \label{scd+} \\
\sigma_{C,d}^- &\sim& \left[ (h_1^{u} + h_1^{d})\,(4\Delta^N D_{dis} + 
\Delta^N D_{fav}) + (h_1^{\bar u} + h_1^{\bar d})\,(\Delta^N D_{dis} + 
4\Delta^N D_{fav}) + 2(h_1^{s} + h_1^{\bar s}) \, \Delta^N D_{1, s}\right] \label{scd-}
\eea
and, dropping the common factor $(1/9)(\alpha^2/Q^4) \, [1+(1-y)^2] A(P_T)$,
\bea
\sigma_{0,p}^+ &\sim& \left[ (4f_1^{u} + f_1^{\bar d})\, 
D_{1,fav} + (4f_1^{\bar u} + f_1^{d})\, D_{1,dis} +
(f_1^{s} + f_1^{\bar s})\, D_{1,s}\right]  \label{s0p+}\\
\sigma_{0,p}^- &\sim& \left[ (4f_1^{u} + f_1^{\bar d})\, 
D_{1,dis} + (4f_1^{\bar u} + f_1^{d})\, D_{1,fav} +
(f_1^{s} + f_1^{\bar s})\, D_{1,s}\right]  \label{s0p-}\\
\sigma_{0,d}^+ &\sim& \left[ (f_1^{u} + f_1^{d})\,(4D_{1,fav} + D_{1,dis}) + 
(f_1^{\bar u} + f_1^{\bar d})\,(D_{1,fav} + 4D_{1,dis}) +
2(f_1^{s} + f_1^{\bar s})\, D_{1,s}\right] \label{s0d+}\\
\sigma_{0,d}^- &\sim& \left[ (f_1^{u} + f_1^{d})\,(4D_{1,dis} + D_{1,fav}) + 
(f_1^{\bar u} + f_1^{\bar d})\,(D_{1,dis} + 4D_{1,fav}) +
2(f_1^{s} + f_1^{\bar s})\, D_{1,s}\right] \>, \label{s0d-}
\eea
where we have used $\sigma_d = \sigma_p+\sigma_n$. 

The suggestion of Ref.~\cite{Barone:2019yvn} is that of measuring ratios of the
difference asymmetries defined in Eq.~(\ref{adt}). This is because, as it can be 
see from the above expressions of $\sigma_C$ and $\sigma_0$, the differences
$(\sigma_{C,p}^+ - \sigma_{C,p}^-)$ and $(\sigma_{C,d}^+ - \sigma_{C,d}^-)$ have 
the same dependence ($\Delta^N D_{fav}-\Delta^N D_{dis})$ on the Collins 
function, which then cancels out in the ratio. Notice that in the differences 
$(\sigma_{C,t}^+ - \sigma_{C,t}^-)$ the contributions from strange quarks drop 
out. Then one has:    
\be
R_{D,d/p} \equiv \dfrac{A_{D,d}}{A_{D,p}} = 3 \left[ \dfrac
{(4f_1^{u} + 4f_1^{\bar u} + f_1^{d} + f_1^{\bar d})\,(D_{1,fav} + D_{1,dis})
+ 2(f_1^{s} + f_1^{\bar s})\,D_{1,s}}
{5(f_1^{u} + f_1^{\bar u} + f_1^{d} + f_1^{\bar d})\,(D_{1,fav} + D_{1,dis})
+ 4(f_1^{s} + f_1^{\bar s})\,D_{1,s}} \right]
\dfrac{h_1^{u_v} + h_1^{d_v}}{4h_1^{u_v} - h_1^{d_v}} \label{ratio}
\ee
as in Eq.~(16) of Ref.~\cite{Barone:2019yvn} and where 
$h_1^{q_v}=h_1^{q}-h_1^{\bar q}$.

Notice that Eq.~(\ref{ratio}) further simplifies if one neglects the contribution
of $s$ quarks. It gives a direct access, assuming one knows the unpolarised
PDFs and FFs, to the ratio of the $u$ and $d$ transversity distributions.

The advantage of suggesting a measurement of $R_{D,d/p}$, apart from its 
direct relation to $h_1^{d_v}/h_1^{u_v}$, is that it can be obtained from 
available data on the Collins asymmetry and the unpolarised cross section
$\sigma_0$. However, it has the disadvantage that it is a ratio of two very 
small quantities, $A_{D,d}$ and $A_{D,p}$, both with large relative errors. 
Then, their ratio is bound to have huge uncertainties, as pointed out also in 
Refs.~\cite{Barone:2019yvn} and~\cite{compass:2020}. The situation might improve 
with the planned next COMPASS run with a deuteron target~\cite{compass:2020}. 
Such a run might offer new possibilities, like the measurement of cross sections. 
In the next Section we suggest another way of obtaining direct information on 
the ratio of transversity distribution for $u$ and $d$ valence quarks.

\section{\label{conc}New measurements, numerical estimates and conclusions}

\begin{figure}[]
	\begin{center}
		\begin{subfigure}{.5\textwidth}
			\includegraphics[width=9cm,height=6cm,clip]{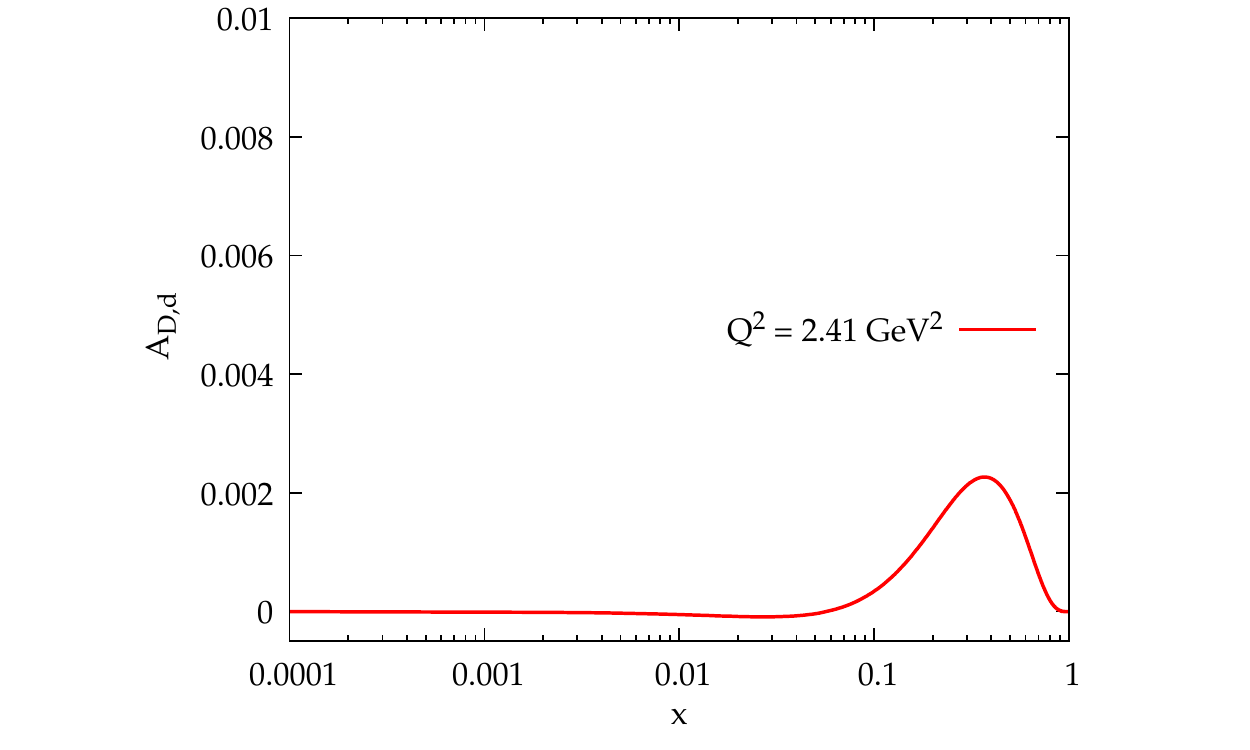}
		\end{subfigure}%
		\begin{subfigure}{.5\textwidth}
			\includegraphics[width=9cm,height=6cm,clip]{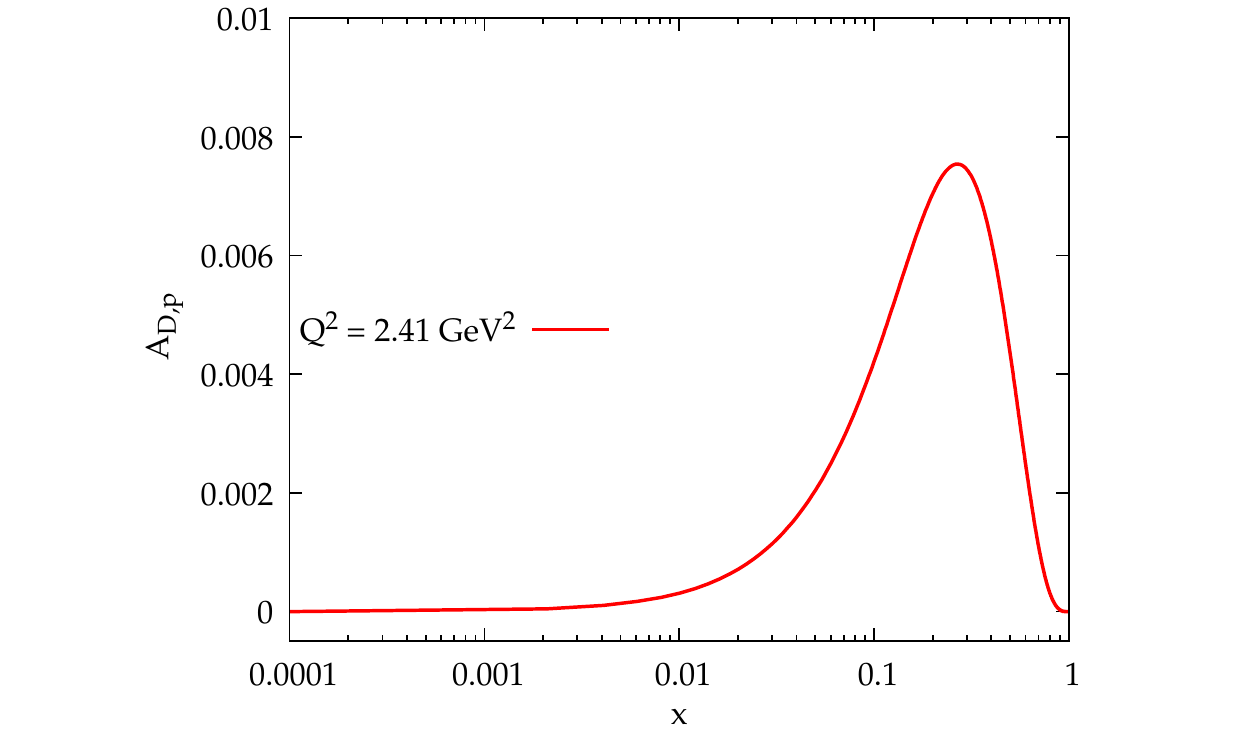}
		\end{subfigure}
	\end{center}
	\caption{\label{fig1} Plot of the difference asymmetries $A_{D,d}$ 
	(left panel) and $A_{D,p}$ (right panel) vs.~$x$ at $Q^2 = 2.41$ GeV$^2$. 
	The $z$ and $P_T$ variables are integrated in the ranges $0.1 < z < 1$ 
	and $0 < P_T < 5$ GeV.}
	\label{figure1}
\end{figure} 

\begin{figure}[]
	\begin{center}
		\begin{subfigure}{.5\textwidth} 
			\includegraphics[width=9cm,height=6cm,clip]
			{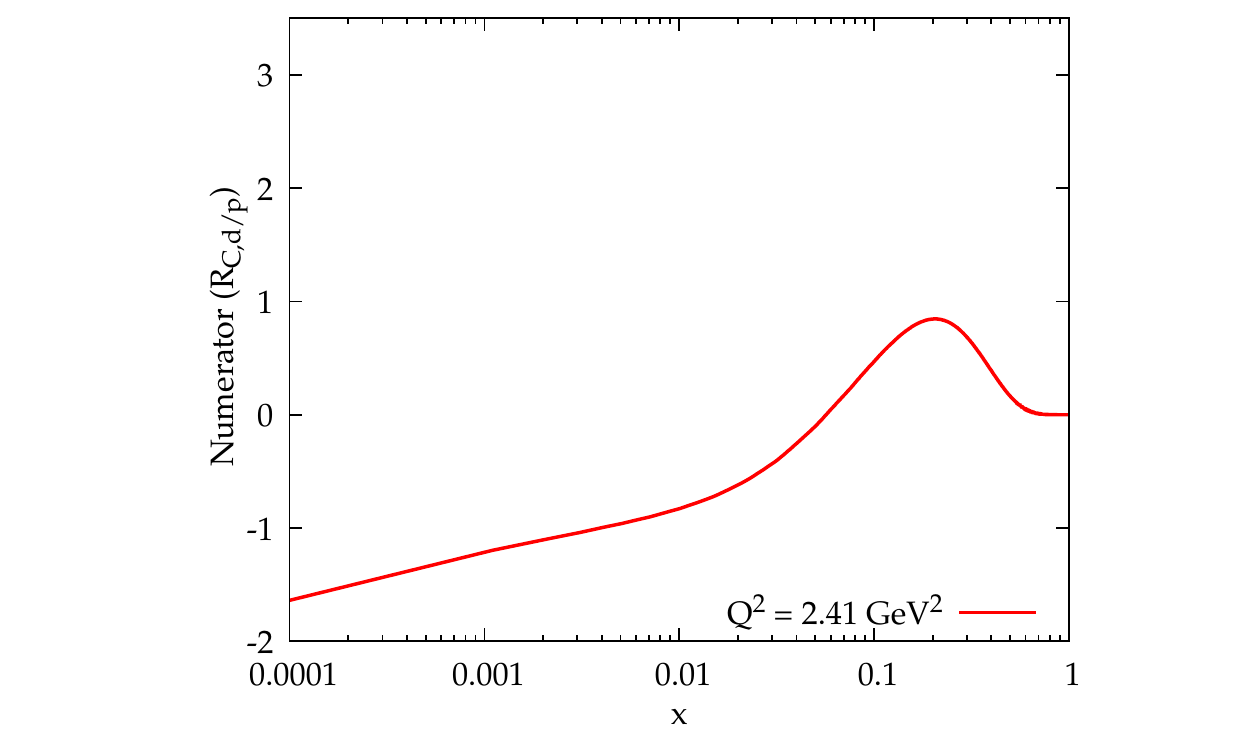}
		\end{subfigure}%
		\begin{subfigure}{.5\textwidth}
			\includegraphics[width=9cm,height=6cm,clip]
			{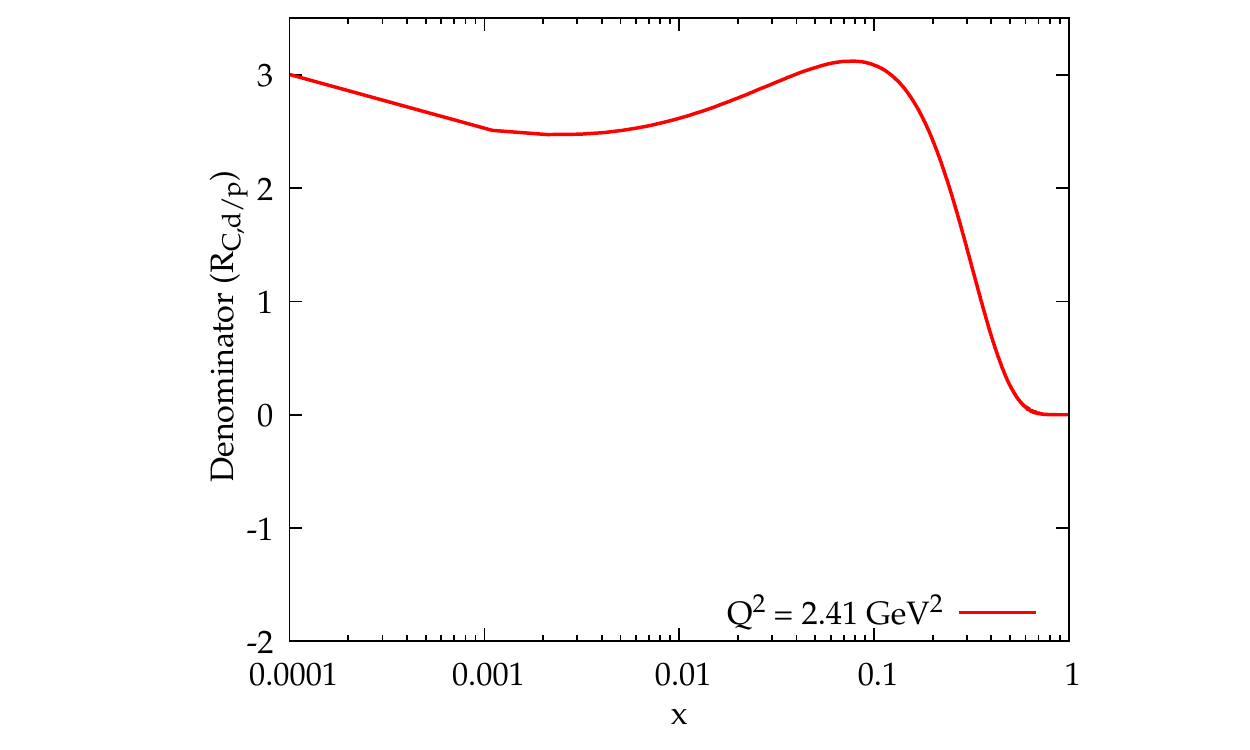}
		\end{subfigure}
	\end{center}
	\caption{\label{fig2} 
	Plot of the numerator (left panel) and denominator (right panel) of 
	$R_{C,d/p}$  vs.~$x$ at $Q^2 = 2.41$. The $z$ variable is integrated 
	in the ranges $0.1 < z < 1$.}
	\label{figure2}
\end{figure}

If one could measure the $\sin(\phi_h + \phi_S)$ modulation of the SIDIS
cross section (\ref{eq:dsig-sidis}), that is $\sigma_C$, for different targets and 
positive and negative pions, then one could built the ratio
\be
R_{C,d/p} \equiv \dfrac{\sigma_{C,d}^+ - \sigma_{C,d}^-}
{\sigma_{C,p}^+ - \sigma_{C,p}^-} \label{ratio2}
\ee   
which, from Eqs.~(\ref{scp+})--(\ref{scd-}), has a very simple partonic
interpretation:
\be
R_{C,d/p} = 3 \, \frac{h_1^{u_v} + h_1^{d_v}}{4h_1^{u_v} - h_1^{d_v}} 
= 3\, \frac{1 + \dfrac{h_1^{d_v}}{h_1^{u_v}}} {4 - \dfrac{h_1^{d_v}}{h_1^{u_v}}}  
\> \cdot\label{newratio}
\ee 
Not only the dependence on the Collins functions cancel out, like in 
Eq.~(\ref{ratio}), but also the dependence on the unpolarised PDFs and FFs.
Moreover, $R_{C,d/p}$, compared to $R_{D,d/p}$, is not a ratio of two small quantities, with a consequent smaller uncertainty. This is essentially due to 
the fact that the $A_D$'s are related to a ratio $\sigma_C/\sigma_0 \sim  
(h_1 \, \Delta^ND)/(f_1 \, D)$, while the numerator and denominator of $R_C$ 
are simply proportional to $\sigma_C \sim (h_1 \, \Delta^ND)$.   

A similarly simple expression holds if one measures the SIDIS cross section 
off a neutron target, possibly at JLab~\cite{Gao:2018grv}:
\be
R_{C,n/p} = \frac{4 h_1^{d_v} - h_1^{u_v}}{4h_1^{u_v} - h_1^{d_v}} 
= \frac{4\dfrac{h_1^{d_v}}{h_1^{u_v}} - 1} {4 - \dfrac{h_1^{d_v}}{h_1^{u_v}}}  
\> \cdot\label{newration}
\ee 

We can give some estimates for the suggested asymmetries, based on our actual 
knowledge of the transversity distributions. We use the simple parameterization 
for the transversity distributions and the Collins functions as in Eqs.~(9)--(12) 
of Ref.~\cite{Anselmino:2013vqa}. In that reference, the best fit free parameters 
for the $u$ and $d$ quarks transversity distributions functions and for the 
favoured and disfavoured Collins fragmentation functions have been extracted 
by fitting HERMES, COMPASS and Belle data. For our plots we use the values 
given in Table~II of Ref.~\cite{Anselmino:2013vqa}. Notice that the transversity 
distributions for $\bar u$ and $\bar d$ are assumed to be negligible.

In Fig.~\ref{figure1} we plot $A_{D,d}(x)$ and $A_{D,p}(x)$ defined in 
Eq.~(\ref{adt}), using Eqs.~(\ref{scp+})--(\ref{s0d-}) where we have 
reinserted all factors. The $y$ dependence cancels out, while the $P_T$ and 
$z$ variables are integrated in the ranges $0 < P_T < 5$ GeV and $0.1 < z < 1$. 
We have fixed $Q^2 = 2.41$ GeV$^2$, which is the $Q^2$ value of the results 
of Ref.~\cite{Anselmino:2013vqa}. The PDFs are taken from 
MSTW2008~\cite{Martin:2009iq} and the unpolarised pion FFs from 
Ref.~\cite{deFlorian:2014xna}; for the helicity distributions we refer to 
Ref.~\cite{Gluck:2000dy}. Very similar results could be obtained simply
using directly in Eq.~(\ref{adt}) the expressions~(\ref{scp+})--(\ref{s0d-}),
and integrating over $z$.  
   
In Fig.~\ref{figure2} we show the numerator and denominator of 
$R_{C,d/p}$, respectively $(\sigma_{C,d}^+ - \sigma_{C,d}^-)$ and 
$(\sigma_{C,p}^+ - \sigma_{C,p}^-)$ as obtained from 
Eqs.(\ref{scp+})--(\ref{scd-}). The $z$ variable is integrated between 0.1 
and 1. A similar plot could be shown for the numerator of $R_{C,n/p}$.  


These estimates clearly confirm 
our expectations. The difference asymmetries, $A_{D,d}(x)$ and $A_{D,p}(x)$,
available from existing data, are, however, very small; their uncertainties,
due to experimental errors and difficulties, can be as large as their values; 
their ratio, which could avoid a knowledge of the Collins function, is bound 
to have huge uncertainties~\cite{Barone:2019yvn, compass:2020}.      

Our suggested measurements of $R_{C,d/p}$ and $R_{C,n/p}$ require a knowledge 
of the SIDIS cross section~(\ref{eq:dsig-sidis}) and in particular of its 
$\sin(\phi_h + \phi_S)$ modulation, which might be difficult. However, their 
TMD interpretation is much cleaner and allows a direct measurement of the 
ratio $h_1^{d_v}/h_1^{u_v}$, through the ratio of two quantities which can be 
orders of magnitude larger than the difference asymmetries.

We are confident that the simplicity of $R_{C,d/p}$ and $R_{C,n/p}$ in terms 
of the transversity distributions, Eqs.~(\ref{newratio}) and ~(\ref{newration}), 
will prompt and encourage their measurements; this could be done during the next 
deuteron COMPASS run, or during the ongoing JLab 12 experiments or at the future 
EIC facility.      


\acknowledgments 
M.A. would like to thank the Physics Department of the Indian Institute of 
Technology in Mumbai (IIT Bombay), where this paper was initiated, for 
hospitality and support. 
We are grateful to F. Bradamante and V. Barone for useful comments and discussions
and to U. D'Alesio for some help with the pion FFs.

\end{document}